\documentclass[%
 reprint,
 amsmath,amssymb,
 aps,
]{revtex4-2}

\usepackage{graphicx}% Include figure files
\usepackage{dcolumn}% Align table columns on decimal point
\usepackage{bm}% bold math
\usepackage{xcolor}
\usepackage{float}

\newcommand\varpm{\mathbin{\vcenter{\hbox{%
  \oalign{\hfil$\scriptstyle+$\hfil\cr
          \noalign{\kern-.3ex}
          $\scriptscriptstyle({-})$\cr}%
}}}}
\newcommand\varmp{\mathbin{\vcenter{\hbox{%
  \oalign{$\scriptstyle({+})$\cr
          \noalign{\kern-.3ex}
          \hfil$\scriptscriptstyle-$\hfil\cr}%
}}}}

\begin{document}

\title{Time-varying media, relativity, and the arrow of time}

\author{Matias Koivurova$^{1,2}$}
\email{matias.koivurova@tuni.fi}
\author{Charles W. Robson$^2$}
\author{Marco Ornigotti$^2$}

\affiliation{$^1$Tampere Institute for Advanced Study, Tampere University, 33100 Tampere, Finland}
\affiliation{$^2$Faculty of Engineering and Natural Sciences, Tampere University, 33720 Tampere, Finland}

\date{\today}

\begin{abstract}
We study the implications of time-varying wave mechanics, and show how the standard wave equation is modified if the speed of a wave is not constant in time. In particular, waves which experience longitudinal acceleration are shown to have clear relativistic properties when a constant reference speed exists. Moreover, the accelerating wave equation admits only solutions propagating forward in time, which are continuous across material interfaces. We then consider the special case of electromagnetic waves, finding that the Abraham-Minkowski controversy is caused by relativistic effects, and the momentum of light is in fact conserved between different media. Furthermore, we show that the accelerating waves conserve energy when the wave is moving along a geodesic and demonstrate two example solutions. We conclude with some remarks on the role of the accelerating wave equation in the context of the arrow of time.
\end{abstract}

\maketitle

\section{Introduction}

Recently, time-varying media has risen as a new paradigm in wave mechanics, in particular for the case of electromagnetic radiation \cite{Review}. This is due to the emergence of highly nonlinear materials, such as epsilon-near-zero (ENZ) media, which may feature large and fast modulation to the dielectric function under specific conditions. Time-varying media have been shown to host a range of interesting effects, such as temporal scattering \cite{time_scatter}, time-crystals \cite{TC1,TC2,TC3}, Photonic time crystals \cite{DPC1,DPC2}, and antireflection temporal coatings \cite{antiref}, to name a few. What has been neglected so far, is that when the properties of matter change, the wave propagating inside it experiences longitudinal acceleration. This is true no matter what type of wave we are discussing.

The speed of light may be the most important constant of nature. It governs the physics of the small and large, sets boundaries in the theory of relativity \cite{Barut}, and is directly linked to a large number of observables \cite{huang, jackson, storyC}. In the framework of modern physics, it is normal practice to take the speed of light as a strictly constant value \cite{Landau&Lifshitz}. However, we encounter situations where it varies every day: whenever light interacts with matter, it appears to slow down. Obviously, this is not a modification to the vacuum speed of light but it is an important property nonetheless.

However, this presents a problem: although electromagnetic waves clearly experience longitudinal acceleration, such accelerating waves are \textit{not} solutions of the standard wave equation, although they are solutions to Maxwell's equations. Motivated by this observation, it is clear that the same problem persists in all wave mechanics, whether electromagnetic or mechanical. In the present work, we examine the implications of a varying speed in wave propagation, and derive a general wave equation for this case. The considerations here apply to propagation through free-space as well as any media. We start by deriving a wave equation for arbitrary accelerating waves, and present a general solution to it. The solutions have a preferred direction of time, therefore establishing the microscopic arrow of time. We then continue by concentrating on the special case of electromagnetic waves, and show that an electromagnetic wave that has a varying speed will experience relativistic effects. Moreover, when the wave is moving along a geodesic, its energy and momentum are conserved. This has important implications on one of he longest standing debates in photonics. We then consider a curved spacetime geometry, which causes the wave to gain energy upon propagation.

\section{The accelerating wave equation}

For smooth functions differentiation is commutative, i.e., the result of iterated derivation does not depend on the order in which the single derivatives are applied. Therefore, multiplication with first order differential operators on a set of smooth functions makes perfect sense in a quite general way \cite{Nakahara}. For the sake of simplicity, in what follows we limit ourselves to consider differential operators on twice differentiable functions, in which case higher order derivatives may not be defined, but the multiplication with differential operators is still valid.

\subsection{Derivation and a general solution}

Let us consider 1+1-dimensional motion for simplicity, with the understanding that the following calculations can be generalized to three spatial coordinates. Let us say that we have a function $f$ which exists on a spacetime grid $(x,t)$ such that $f\equiv f(x,t)$. With the use of the triple product rule, we can write \cite{thermo}
\begin{align}
    \left.\left( \frac{\partial f}{\partial t}\right)\right\rvert_x
    \left.\left( \frac{\partial x}{\partial f}\right)\right\rvert_t
    \left.\left( \frac{\partial t}{\partial x}\right)\right\rvert_f 
    = -1.
    \label{triple}
\end{align}
Since we are considering motion, the third partial derivative can be interpreted as the inverse speed of the function $f$ on the grid, that is
\begin{align}
    \left.\left( \frac{\partial x}{\partial t}\right)\right\rvert_f = c,
\end{align}
which we may plug into Eq.~(\ref{triple}) without loss of generality.

Furthermore, if we assume that the function $g = \partial x/\partial f$ is a smooth enough function on a given domain $D = \{g \in C^1: g(f) \neq 0 \}$, then it is well-behaved, and in particular, its inverse, $g^{-1} = \partial f/\partial x$, exists and is well-behaved as well. We may then multiply both sides of Eq.~(\ref{triple}) with $c g^{-1}$, and collect everything on the l.h.s to obtain
\begin{equation}
    \left( \frac{\partial}{\partial t} + c\frac{\partial}{\partial x}\right) f(x,t) = 0,
    \label{one-way}
\end{equation}
where we have left the vertical evaluation lines implicit for brevity. Furthermore, if we multiply both sides with $\partial/\partial t - c \partial/\partial x$, and assume that $c$ is constant, we arrive at the familiar wave equation
\begin{equation}
     \frac{\partial^2f(x,t)}{\partial t^2} = c^2 \frac{\partial^2f(x,t)}{\partial x^2},
\end{equation}
revealing that the function $f$ is some type of wave. In earlier works, this last step has been done the other way around to obtain a one-way wave equation (Eq.~(\ref{one-way}), see e.g. Ref.~\cite{one_way}).

The reader should keep in mind that so far, the derivation presented above does not offer any information on the nature of the field $f$, and such information has to be inferred from somewhere else (e.g., from Maxwell's equations). Remarkably, the only assumption that is required to arrive at the standard wave equation is that the speed of the wave is constant everywhere and at all times. 

Let us now consider the case, in which we allow the speed of the wave to vary with time, i.e., we set $c \equiv c(t)$. The above treatment then yields a wave equation for an accelerating wave
\begin{equation}
    \frac{\partial^2f(x,t)}{\partial t^2} = c(t)^2 \frac{\partial^2f(x,t)}{\partial x^2} - \dot{c}(t) \frac{\partial f(x,t)}{\partial x},
    \label{new}
\end{equation}
where $\dot{c} = \partial c/ \partial t$ is the acceleration of the wave $f$. Note that there is no explicit dependence on position, $x$, and a time dependence is adequate for quantifying the spatial variation of propagating waves.

Let us now analyze Eq.~(\ref{new}). If we take the spatial Fourier transform of the wave, defined as $\tilde{f}(k,t) = \int_{-\infty}^\infty f(x,t) \exp(ikx) dx$, where $k = \omega/c_0$ is the wave number, $\omega$ is the angular frequency, and $c_0$ is some constant reference speed, we attain an equation of the form
\begin{equation}
    \frac{\partial^2\tilde{f}(k,t)}{\partial t^2} = -\Omega^2(t) \tilde{f}(k,t),
    \label{oscillator}
\end{equation}
where $\Omega^2(t) = k^2c(t)^2 - ik\dot{c}(t)$. This equation is analogous to a harmonic oscillator with a time dependent frequency, and analyzing it usually requires the use of invariants or other sophisticated mathematical tools \cite{invariant1, invariant2, invariant3, path_int}. However, in the present case where $\Omega(t)$ depends on the speed and acceleration of the wave, there is a simple plane wave solution of the form
\begin{align}
    \tilde{f}(k,t) \propto \exp \left[ \varpm \, i \frac{\omega}{c_0} \int c(t) dt \right],
    \label{solution}
\end{align}
where the minus sign is allowed only when $\int c(t) dt<0$ (or $\omega<0$). That is, the sign of the solution has to be positive. This can be seen by inserting Eq.~(\ref{solution}) into Eq.~(\ref{oscillator}).

Intriguingly, we can now draw direct parallels to the theory of relativity by choosing the reference speed, $c_0$, as the vacuum speed of light, and writing the speed variations as a modulation to this speed such that $c(t) = c_0 n(t)$, where $n(t) \in [0,1]$. With this choice, we immediately identify the integral in Eq.~\eqref{solution} to be analogous to the \textit{proper time} found in general relativity \cite{Landau&Lifshitz}. Namely $t'=\int \sqrt{g_{tt}} \, dt$, where $g_{tt}$ is the time component of the metric $g_{\mu\nu}$ characterising the curved spacetime in which the time experienced by the wave is calculated. Therefore, $n(t) \equiv \sqrt{g_{tt}}$, and we can identify the time $t$ as the \textit{coordinate time} measured by a stationary observer. The ratio between these two times is what defines \textit{time dilation}.

Going back to the spatial domain, we can write the general solution as a superposition of waves moving towards the positive and negative directions along the space axis,
\begin{align}
    f(x,t) = A\exp(i\omega t' - ikx) + B\exp(i\omega t' + ikx),
    \label{solution_space}
\end{align}
where the amplitudes $A$ and $B$ depend on the particular solution. Let us briefly consider the spatial coordinates in the different frames with a constant $n(t) = n$ for simplicity. In general, a constant $n$ defines a geodesic in a Minkowski-type spacetime with mostly positive signature \cite{Barut}. In the stationary frame we have $dx/dt = c(t)$ such that the leading edge of the wave is at $x = \int c(t) dt = c_0 t'$, where $t'=tn$. If we multiply both sides by $k$, we find that $kx = \omega t'$. For a stationary observer who describes the dynamics with respect to the laboratory time, the relation $kx' = \omega t$ must hold, which leads to $x' = x/n$. In other words, the wave will undergo a \textit{length contraction} in the stationary frame. From here, it is easy to see that when there are no gravitational effects, the function $n(t)$ reduces to the Lorentz factor $\gamma(t) = [1-c(t)/c_0]^{-1/2}$ of special relativity. 

The above reasoning shows that not only matter, but also waves experience relativistic effects. In particular, a wave that is moving at a significant fraction of the speed of light will experience special relativistic effects, whereas a longitudinally accelerating wave experiences general relativistic effects, reminiscent of gravitation. It needs to be stressed that these considerations apply to all waves, whether they are electromagnetic or mechanical in nature.

\subsection{The arrow of time}

Let us briefly review our understanding of time. In everyday experience, time always seems to have a preferred direction. That is, time always moves towards the future and away from the past. However, there is no concrete reason why this should be. As Sir Arthur Eddington wrote \cite{Eddington}: ``Let us draw an arrow arbitrarily. If as we follow the arrow we find more and more of the random element in the state of the world, then the arrow is pointed towards the future; if the random element decreases, the arrow points towards the past. That is the only distinction known to physics.'' This was how Eddington coined the term ``arrow of time,'' and tied it explicitly to the second law of thermodynamics. Up to now, this has been the best explanation for the existence of a preferred direction of time. Nonetheless, it is only an observation of the behaviour of macroscopic objects, and not an actual law of nature.

\begin{figure}[ht!]
    \centering
    \includegraphics[width=\columnwidth]{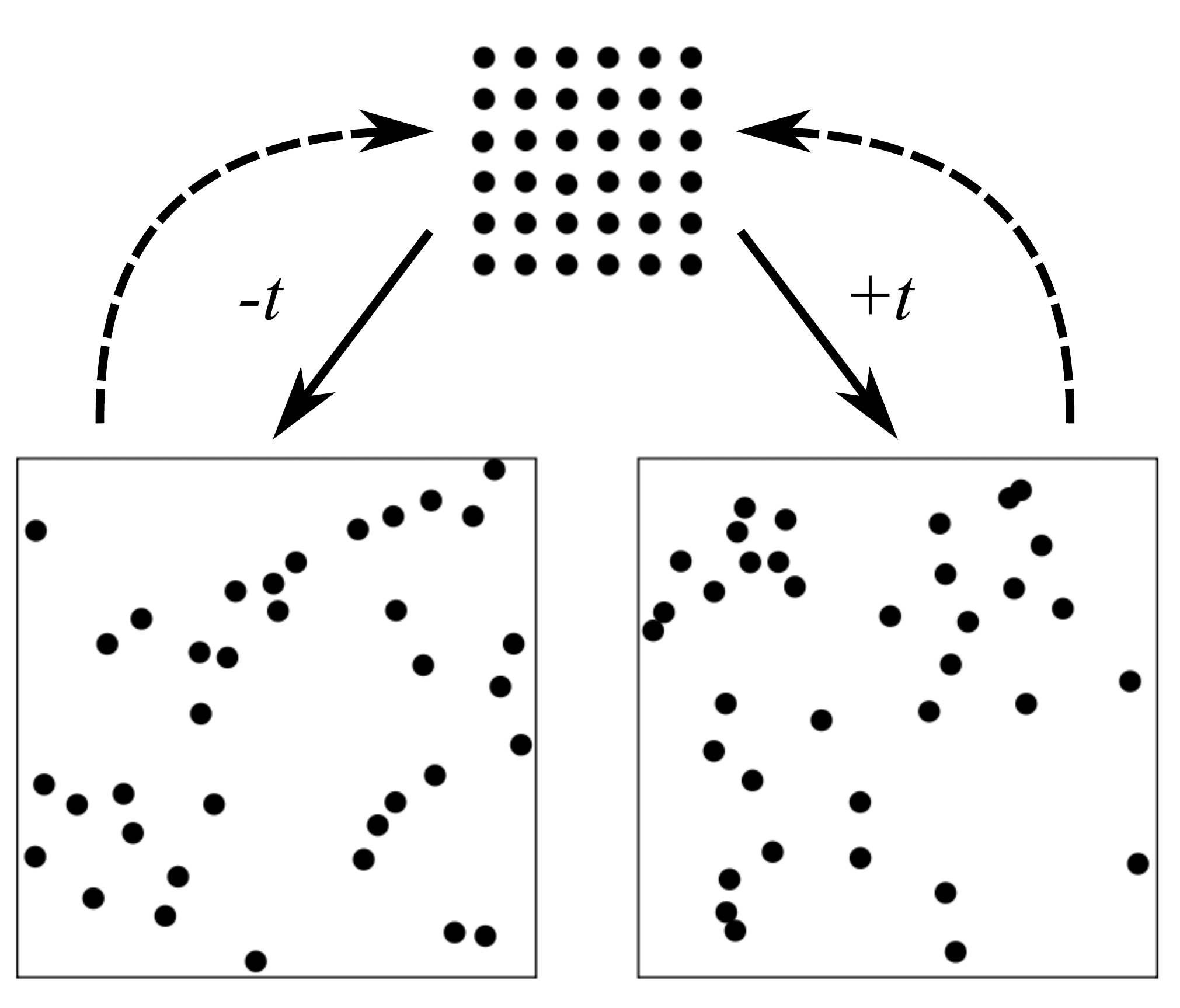}
    \caption{Visualization of a collection of repulsive particles in a box. Disorder (and thus entropy) increases whether the microscopic arrow of time points towards positive \textit{or} negative time. The dashed arrow depicts a time reversal, causing disorder to decrease until the particles reach their initial configuration.}
    \label{particles}
\end{figure}

To see why this is, consider a collection of repulsive particles in a highly ordered configuration and inside a bounded region, which then evolves toward \textit{either} positive \textit{or} negative time.  This situation is depicted schematically in Fig.~\ref{particles}. In both cases, the total entropy will increase. However, if time were to be reversed at any point during the evolution (dashed lines in Fig.~\ref{particles}), entropy would decrease until the particles go back to their initial highly ordered configuration, and then start rising again. Therefore, increasing entropy is a sign of macroscopic arrow of time, and it doesn't tell us anything about the direction of microscopic time (Ref.~\cite{Eddington}, p. 79), or why time cannot be reversed (Ref.~\cite{Eddington}, p. 93--94).
 
Here, we shed light on these questions. An observation of great importance is that when a wave experiences longitudinal acceleration, Eq.~(\ref{new}) allows only solutions that propagate \textbf{towards positive time}, as seen from  Eq.~(\ref{solution}). Moreover, reflection along temporal axis (i.e., true time reversal) is allowed only when $\dot{c}(t) = 0$, but any reflection process must induce an acceleration. Thus, a wave must experience a positive time, which cannot be reversed. Note that this is distinct from time reflection discussed e.g. in Refs.~\cite{1,2,3,4,5,6} in the context of electromagnetic time dependent media. This ambiguity in terminology requires some consideration.

In the context of electromagnetic waves, time dependent media has been studied at least since the 1950s \cite{time_scatter,Review}, but the term time refraction has been introduced only relatively recently \cite{1,2}. Multiple studies on spatiotemporal effects can be found in existing literature, where the term time reflection or time refraction is often used explicitly \cite{1,2,3,4}. However, this usually does not mean a reflection or refraction along the time dimension, but rather, a spatial reflection $x \rightarrow -x$, due to a time dependent change in the material properties. For example, in Refs.~\cite{2,3}, the case of an abrupt and uniform change of the physical properties of a bulk material were investigated, and the properties of time reflection and refraction were discussed in detail.

This is the most common use of the term ``time reflection,'' and it is entirely valid in the form it has been introduced in the literature. In the present study, we discuss reflection as a mathematical process \textit{along} the temporal dimension with respect to the origin of time, i.e. $t \rightarrow -t$, while leaving the spatial coordinate unchanged. This constitutes a \textit{true time reversal} for an arbitrary wave, which is an entirely different concept. Indeed, if the time of a wave could be exactly reversed, it would propagate in a manner that allows it to reverse it's direction and even any diffraction effects.

Further, it needs to be noted that studies on time reversal have been successfully carried out with optical beams \cite{5}. This has been realized with Pendry's four-wave mixing method \cite{6}, where a transition of $-2\omega$ is induced on a probe beam of frequency $\omega$ to cause a shift to $-\omega$, i.e. a reflection along the frequency axis $\omega \rightarrow -\omega$. Since a plane-wave has a time dependence of the form $\exp(i\omega t)$, this has a similar effect as reversing time. Our theoretical results are not contradicting this in any way, since even in this type of experiment time still flows forward.

The time asymmetry of the solutions we report is in fact the first theoretical result on the \textit{microscopic} arrow of time. In other words, we have found a plausible explanation as to why time has a preferred direction, at least in terms of wave mechanics. Moreover, we show that there is a plausible theoretical reason why time cannot be reversed, and thus entropy must increase in all cases, thus constituting an important addendum to the macroscopic arrow of time. Since the accelerating wave equation is universal, this appears to be widely applicable.

\section{Accelerating electromagnetic waves}

To gain further insight into the accelerating wave equation and its solutions, we need to choose the type of wave to be studied. Electromagnetic fields constitute an important special case of longitudinally accelerating waves, since they travel at the vacuum speed of light by default, and we can only decelerate them. This leads to some unintuitive properties, which we will be discussed in length. In the case of electromagnetic radiation, $c(t)$ is the speed of light in the above equations, and $f$ is the electric (or magnetic) field.  For electromagnetic waves the modulation to the vacuum speed is more naturally written as $c(t) = c_0 / n(t)$, so that $n(t) \in [1,\infty]$ and it can be interpreted as a refractive index. In principle, lower values of $n$ are also possible, but such cases require dispersion for them to be physical.

%Note that an alternative derivation for Eq.~(\ref{new}) would be to use Fa\`a di Bruno's formula \cite{Porteus} on the second derivative of the electric field, and assume that the speed of light is not constant.
 
Now, we can define an \emph{intrinsic} time for the electric field $f(x,t)$, as $t' = \int n(t)^{-1} dt$, which is a time experienced by the electromagnetic wave. The notion of time in the frame of an electromagnetic wave may seem faulty, since photons move along null geodesics. Still, intrinsic time is well-defined, since electromagnetic waves still accumulate a phase on propagation, which depends on time. Notice, how in general a constant $n(t)=n$ defines a geodesic in a Minkowski-type spacetime with mostly positive signature \cite{Barut}, but is characterised by an effective speed of light $c=c_0/n$. The usual Minkowski spacetime with $c=c_0$ is then obtained for the particular value of $n=1$, i.e., when the wave is propagating in vacuum. In other words, the acceleration $\dot{c}(t)$ effectively induces a nonstationary metric $g_{\mu\nu}$ in the moving reference frame of the electric field. In such a frame, the dynamics of the field $f(x,t)$ are that of a field evolving on a curved background. However, an external observer will measure a time $t$ and describe the dynamics of $f(x,t)$ in a flat spacetime, as usual.

If we again consider the leading edge of the wave at $x = c_0 t'$, where the intrinsic time is $t' = t/n$, we find that the electromagnetic wave experiences \textit{time contraction} instead of time dilation. This follows from the fact that we are decelerating the wave from it's usual speed, as opposed to the acceleration of mechanical waves considered above. Moving on in the same manner as in the mechanical case, we multiply both sides by $k$ to get $kx = \omega t'$ and for a stationary observer who describes the dynamics with respect to the laboratory time, we get $x' = xn$ so that the wave undergoes \textit{length dilation}. Again, this is entirely due to the fact that we are causing the wave to slow down, and a positive acceleration would cause the usual length contraction. However, this effect is usually absorbed into the wavenumber, and thus we can define $k' = k n$. This is exactly the relation that is employed in conventional optics and photonics.

Note that when the speed of the wave from an outside observer's point of view is equal to the vacuum speed, the intrinsic time corresponds to the laboratory time, and the acceleration term in Eq.~(\ref{new}) is null, so that the above reasoning reduces to the usual wave equation and a plane wave solution. 

\subsection{Energy and momentum conservation}

Next, we briefly consider the energy and momentum carried by a longitudinally accelerating wave, as seen by a stationary observer. We can find the energy flux with the Poynting vector $\textbf{S}$, and the average energy density simply by dividing $S = |\textbf{S}|$ with the speed of the wave. Furthermore, the overall energy $E$ can be attained by integrating the average energy density over the whole volume occupied by the wave, which in this case would be of the form
\begin{align}
    E = w\int_{x_0}^{x_1} \frac{S}{c(t)} dx,
    \label{energy}
\end{align}
where $w$ is a constant that comes from integrating along the transverse dimensions, and the coordinates $(x_0, x_1)$ define the length of the volume. The two spatial coordinates can be expressed in terms of intrinsic time as $x_0 = c_0t'_0$, and $x_1 = c_0 t'_1$. Therefore, performing the integral yields $E = w S n(t) \Delta t'$, where $\Delta t' = t'_1 - t'_0$ can be chosen to be a constant. If $n(t)$ defines a geodesic, the above relation yields $E = wS \Delta t$, for any constant value of $n$, and thus, energy is conserved. However, this is generally not the case in a curved spacetime, which is a well-known result in general relativity \cite{gravity1}.

Since it is well-known that the momentum of an electromagnetic wave is found by dividing its energy with its phase velocity, conservation of energy immediately implies conservation of momentum. We can see this also from our plane wave solution: no matter what type of longitudinal acceleration the wave experiences, the wavenumber $k$ remains constant in the moving reference frame of the wave. Since the wavenumber is directly related to the momentum the wave carries, we can conclude that the momentum of a longitudinally accelerating wave is a conserved quantity. More precisely, it is a Lorentz covariant scalar, and thus remains the same under Lorentz transformations.

\subsection{Abraham--Minkowski controversy}

Momentum conservation has a direct consequence on one of the longest standing fundamental questions in photonics: what is the momentum of a photon that enters a dielectric medium? Historically, there have been two contending views, and as such it has been named the Abraham--Minkowski controversy \cite{abraham, minkowski, barnett, contr1,contr2}.

\begin{figure}[ht!]
\centering
\includegraphics[width=\columnwidth]{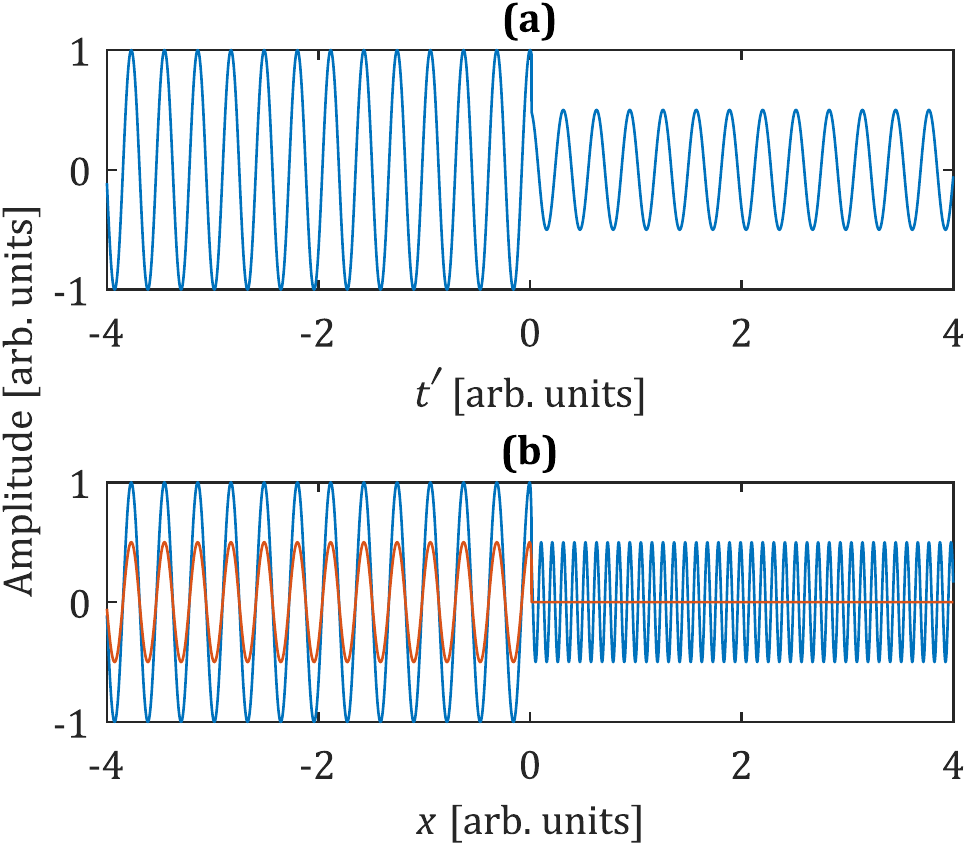}
\caption{Plane wave solution to the accelerating wave equation, (a) in the temporal frame of the wave and (b) in the spatial reference frame of a stationary observer. Blue corresponds to incident and red to reflected wave.}
\label{example1}
\end{figure}

Let us start by applying our plane wave solution to an electromagnetic wave crossing a dielectric interface. Here, the speed of light abruptly changes from some initial value, $c_0$, to something else at time $t=0$ and we can write
\begin{align}
    n(t)^{-1} = \frac{1}{n_1} + \frac{n_1-1}{n_1[1+\exp(2 t m)]},
\end{align}
where $m$ is a free parameter. This is essentially a smooth approximation of the Heaviside step function, when $m$ is sufficiently large. Here $n_1$ is a proportionality factor by which the speed of light changes after $t=0$, i.e. $c(t>0) = c_0/n_1$, and we take $n_1$ to be equal to the refractive index of the material. Integrating with respect to $t$, we obtain the intrinsic time as
\begin{align}
    t' = \frac{t}{n_1} - \frac{n_1-1}{n_1} \frac{\ln[1 + \exp(-2 t m)]}{2 m}.
\end{align}
This ensures that before $t = 0$, the time of the wave will flow as in free space ($t' = t$) whereas after entering the medium, intrinsic time becomes $t' = t/n_1$. A depiction of the above scenario is shown in Fig.~\ref{example1}, for $\omega = 20$, $n_1=3$, $c_0=1$, and amplitude coefficients are calculated with Fresnel equations. The reflected wave does not appear in the temporal reference frame for clarity.

As can be seen from Fig.~\ref{example1}, our approach with a smooth, continuous electric field crossing a dielectric boundary is equivalent to the standard formulation of a piecewise continuous electric field defined via electromagnetic boundary conditions (at least in this simple case). In fact, it would appear that we can reformulate many of the problems in optics and photonics in terms of accelerating waves and do away with dielectric boundaries. Moreover, the interpretation of the wave properties becomes more straightforward within the present framework; if we insist on the intrinsic time description, both energy and momentum are conserved since $\omega$ and $k$ are both constant. If we instead say that time should always be taken in the outside observer's reference frame, then the variation can be transferred to the frequency, such that $\omega' = \omega/ n_1$. If we insist on this explanation, then we are forced to choose a new wavelength inside the medium to conserve energy. However, if we do that then momentum is no longer conserved, leading to the famous Abraham-Minkowski controversy.

Here, we argue that the whole controversy stems from relativistic effects, and is in fact, not a problem at all when inspected in the correct frame of reference. Indeed, one can make compelling arguments for both $n_1\lambda$ and $\lambda/n_1$ cases that are equally correct. However, neither reaches the root of the problem, which is that the apparent change in the wavelength (and thus momentum) is a relativistic effect, and the momentum of the wave is in fact unchanged. Our findings are supported by the heuristic arguments recently laid out in Ref.~\cite{heuristic}. In contrast to earlier resolutions to the controversy, the present formalism does not require a division between material and optical momenta, greatly simplifying the physical picture.

\subsection{Time-varying media}

As a second example of the present study, we look at disordered photonic time crystals \cite{DPC1,DPC2}, which is a prime example of a time-varying medium. We examine this special case, since the physical situation is governed by the accelerating wave equation. A disordered photonic time crystal is defined as a medium where the dielectric function changes abruptly and randomly as a function of time. So far, the photonic time crystal has been studied with the use of discrete time modulations to the dielectric constant. In Ref.~\cite{DPC1} the authors demonstrated via numerical simulations that when the modulation to the dielectric function is fast and strong enough, then a pulse propagating inside a disordered photonic time crystal features an exponentially decreasing speed, while the overall energy of the pulse increases exponentially. This was further corroborated in Ref.~\cite{DPC2}, and this is also the case we shall consider next.

As is well-known, the microscopic explanation for the slowing down of an electromagnetic wave's phase velocity inside a material is due to the disturbance in the charges of each atom. As the electromagnetic field oscillates inside a material, the charges will be accelerated back and forth at the same frequency. This acceleration causes the charges to radiate their own electromagnetic wave at the same frequency, but usually with a phase delay, as the charges may move out of phase with the driving wave. The total electromagnetic wave propagating inside the medium is the superposition of all the waves inside the medium, and since some are out of phase with the driving wave, it will cause the total wave to slow down. This is directly quantified with the refractive index.

Similarly, in a photonic time crystal each time the wave experiences a change in the dielectric function, it will be reflected and refracted. If the modulation of the dielectric function is fast enough, the waves will overlap. The total wave is then again the superposition of all waves present in the material. Continuing to temporally modulate the dielectric function quickly enough will cause an ever increasing number of waves (that are generally out of phase) to overlap inside the material. This will then cause the total wave to slow down, which is accompanied by an increase of amplitude. The effect of a disordered photonic time crystal can therefore be thought to result from the increase of an effective refractive index.

In our study, we do not consider the underlying mechanism that causes the wave to slow down exponentially. Instead, we simply suppose that there is some material where this occurs, and then show how the solutions to the accelerating wave equation behave in this situation. Our analytical solutions would allow us to model the case of discrete modulation as well, which would just be a superposition of a large number of waves similar to the one presented in Fig.~\ref{example1}. While applicable to the discrete case, our results can be thought of as the continuous limit where the length of each modulation segment, $L$, is taken to be vanishingly small, i.e. $L \rightarrow 0$. Indeed, our approach allows for continuous solutions to problems where discontinuous (or piecewise continuous) methods have been used before, such as when light is incident at the interface of two media.

Let us then take $c(t)$ to be of the exponentially decaying form such that $c(t) = c_0\exp(-t/\tau) $, where $\tau$ is the decay half-life. For this particular case, we can write a plane wave solution in the form
\begin{align}
    f(t) =  \exp \left[ i\omega \tau\exp\left(-\frac{t}{\tau}\right)\right],
    \label{crystal}
\end{align}
where we have left the spatial dependence implicit for brevity. Note that we have chosen the minus sign of Eq.~(\ref{solution}), because the intrinsic time of the wave has a negative sign in this case. Employing Eq.~(\ref{energy}) we find that $E \propto \exp(t/\tau)$. In other words, the curved spacetime will cause the wave to violate energy conservation, as expected.

\begin{figure}[ht!]
\centering
\includegraphics[width=\columnwidth]{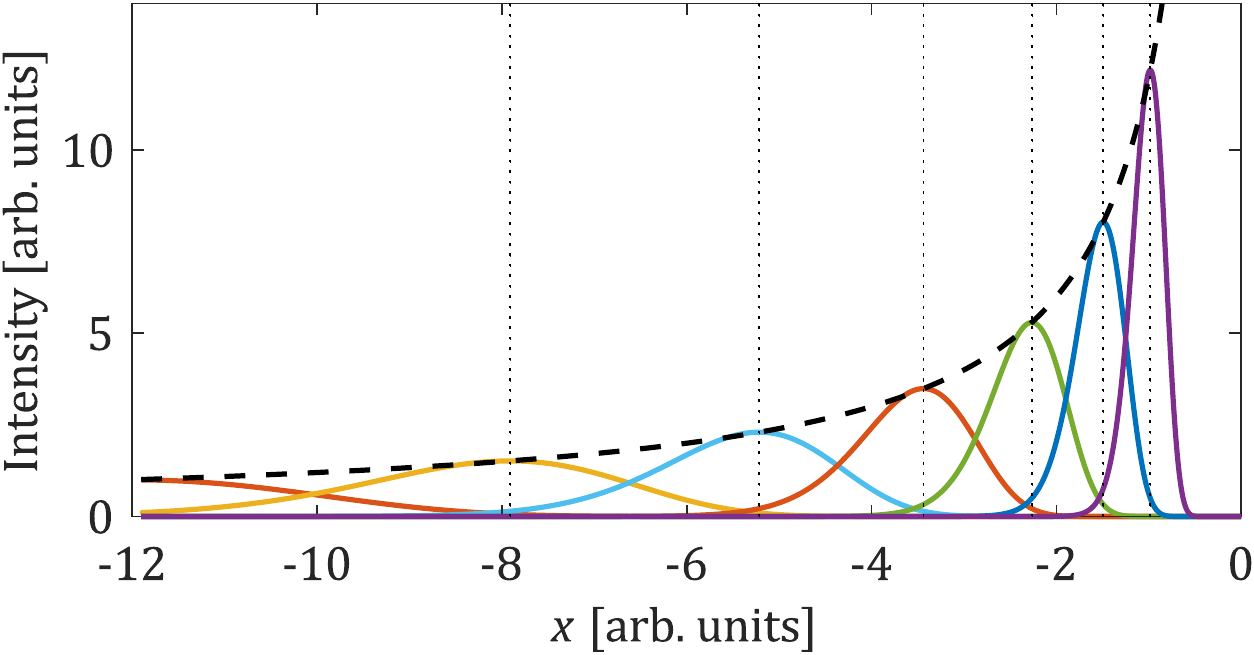}
\caption{Pulse solution to the accelerating wave equation from an outside observer's frame of reference, corresponding to an exponential decrease in speed of the wave. The dashed line shows the evolution of the amplitude, whereas the dotted lines mark the peak position at regular time intervals. The pulse will become shorter as it propagates towards the positive spatial direction and it will never reach $x=0$, because it would require an infinite propagation time.}
\label{example2}
\end{figure}

Since the accelerating wave equation is linear, we can write a superposition integral in the form
\begin{align}
    f(t) = & \int_{-\infty}^\infty \frac{1}{\sqrt{2\pi}} \exp\left[- \frac{(\omega-\omega_0)^2}{2\Omega^2} \right] \nonumber \\
    & \times \exp\left[ i\omega \tau\exp\left(-\frac{t}{\tau}\right)\right] \exp(-i\omega t_0) d\omega,
    \label{crystal_pulse_spec}
\end{align}
which corresponds to a Gaussian pulse with spectral width $\Omega$, and central frequency $\omega_0$, that is positioned at $t_0$ in the temporal domain. Computing the integral yields a temporal domain pulse of the form
\begin{align}
    f(t) = \Omega \exp\left[- \frac{\left(t'-t'_0\right)^2}{2T'^2} \right] \exp\left( i\omega_0't' \right),
    \label{crystal_pulse_temp}
\end{align}
where we have the following exponentially scaled pulse parameters: $T' = \exp(t/2\tau)/\Omega$, $t' = \tau\exp(-t/2\tau)$, $\omega_0' = \omega_0\exp(-t/2\tau)$ and $t'_0 = t_0\exp(t/2\tau)$. In other words, Eq.~(\ref{crystal_pulse_spec}) defines a Fourier transform from spectral domain $\omega$ to an exponentially scaled temporal domain $t'$. Further, Eq.~(\ref{crystal_pulse_temp}) describes an initially Gaussian pulse in a co-moving reference frame. As the pulse moves towards positive time, the amplitude, peak position, width and central frequency all change in unison, although this change is not seen by the wave itself. Figure~\ref{example2} shows the intensity of the pulse solution as seen in the outside observer's spatial frame, with $\tau=12$, $T=3$ and an initial amplitude of unity.

As can be seen from Fig.~\ref{example2}, the pulse slows down, increases in energy, and becomes shorter. Apart from the shortening of the pulse, these results correspond to earlier works. Therefore, it would seem that disordered photonic time crystals are potential analogue systems for studying wave dynamics in a curved spacetime. This is particularly interesting from an experimental point of view, since it has been proposed that such crystals can be created in a laboratory environment, by illuminating an ENZ medium with ultra-short pulses \cite{DPC1}.

Although the present work is entirely theoretical, it is of great interest to consider how such experiments could be realistically carried out. Obviously, the first example in the main text is entirely trivial, as it corresponds to a simple dielectric interface. However, the second example is not as straightforward.

As it was discussed in the supplementary material of Ref.\cite{DPC1}, it is possible to greatly simplify this problem by considering radio waves. In this regime, the frequency as well as the large temporal period make it relatively simple to temporally modulate a signal such that these effects are seen also experimentally. Such experiments have been carried out by modulating the capacitance of transmission lines. Since radio waves and light are both electromagnetic waves, it is to be expected that results with such a simplified system are directly applicable to optical frequencies. However, the experimental implementation is necessarily more complicated.

Again, as it was outlined in the supplementary material of Ref.\cite{DPC1}, experiments in the optical regime are based on nonlinear epsilon near zero (ENZ) materials. ENZ materials are materials with a real part of the permittivity that is close to zero. Due to their strong nonlinear response, the properties of such materials can be greatly modified by illuminating them with ultrashort optical pulses at frequencies sufficiently higher than their plasma frequency. In such a case, an ultrashort pump pulse is absorbed, exciting a large number of electrons to the conduction band, thus modifying the free charge density. At the ENZ wavelength, this causes changes to the refractive index that can be on the order of unity. Since the response time of these materials is on the order of a few femtoseconds, it is then only a matter of structuring the pump pulses in a suitable manner, such that they induce suitable fluctuations to the refractive index which are necessary for experimentally demonstrating the disordered photonic time crystals.

Although, the imaginary part of the dielectric function may cause considerable losses, such that the ENZ material has to be optically thin for observable effects to appear. Moreover, the ultrashort pulses may cause considerable optical damage to the ENZ material, thus limiting the applicability of this approach. Nonetheless, nonlinear ENZ materials are the most promising candidates for experimentally realizing disordered photonic time crystals.

\section{Discussion and conclusions}

To summarise, in this work we have derived a wave equation for longitudinally accelerating waves. The novel accelerating wave equation is applicable to mechanical, as well as electromagnetic waves, and could be of interest to the wider physics community. We then concentrated on electromagnetic waves and, in particular, situations where the speed of light is modulated away from it's vacuum value. In the cases we consider, this modulation corresponds to either a simple change in refractive index, or a disordered photonic time crystal where the speed of light decreases exponentially. We do not consider instances where the speed of light is greater than the vacuum value. Within the analysis we find several important features of accelerating waves. First, the solutions to the accelerating wave equation are continuous everywhere. This is in stark contrast to the usual formulation of photonics, where material interfaces have to be treated via continuity conditions. Second, such waves experience relativistic effects, with a clear connection to the Abraham-Minkowski controversy: if there is a constant reference speed $c_0$, the energy and momentum of the wave is conserved when the wave is moving along a geodesic (even across material interfaces). Third, the exponential increase in the energy of a pulse propagating in a disordered photonic time crystal is due to a curved spacetime that the pulse experiences, and may offer experimental tests of general relativity.

Lastly, the accelerating wave equation allows only solutions that propagate towards \textit{positive} time. This is a very peculiar asymmetry, which has been encountered before only in the context of the weak force \cite{Peskin}. This may have important implications for the arrow of time: since the accelerating wave equation is general, all light should obey it. This may even include force carrier photons, which are not described by any known wave equation. If this is indeed the case, then forces that are mediated via virtual photons \textit{cannot} be time reversed, and they have a preferred direction of time. It needs to be emphasized that this is a novel result, distinct from the thermodynamic arrow of time.

There are several possible future research directions. It would be desirable to generalize the accelerating wave equation to 3 + 1 dimensions, and characterize its solutions, which may be useful in simulation problems. Furthermore, there are experimental tests that can be carried out to confirm some of the findings presented here, most notably in disordered photonic time crystals. By engineering the curved spacetime properly, it may be possible to generate short and intense pulses. Moreover, it would be of interest to see how light behaves in the present formalism as it approaches the event horizon of a black hole.

\begin{acknowledgements}
The authors acknowledge the financial support of the Academy of Finland Flagship Programme (PREIN - decision 320165).
\end{acknowledgements}

\end{document}